\documentclass[twocolumn,showpacs,preprintnumbers,amsmath,amssymb]{revtex4}

\usepackage{graphicx}
\usepackage{dcolumn}
\usepackage{bm}

\begin{document}


\title{Electrical transport in two dimensional electron and hole gas on Si(001)-(2$\times$1) surface}
\author{Hassan Raza}
 \affiliation{School of Electrical and Computer Engineering, Cornell University, Ithaca, NY 14853}%
\author{Tehseen Z. Raza}
 \affiliation{School of Electrical and Computer Engineering, Purdue University, West Lafayette, IN 47907}
\author{Edwin C. Kan}
 \affiliation{School of Electrical and Computer Engineering, Cornell University, Ithaca, NY 14853}

\begin{abstract}
Si(001)-(2$\times$1) surface is one of the many two-dimensional systems of scientific and applied interest. It has two surface state bands (1) anti-bonding $\pi^*$ band, which has acceptor states and (2) bonding $\pi$ band, which has donor states. Due to its asymmetric dimer reconstruction, transport through this surface can be considered in two distinct directions, \textit{i.e.} along and perpendicular to the paired dimer rows. We calculate the zero bias conductance of these surface states under flat-band condition and find that conduction along the dimer row direction is significant due to strong orbital hybridization. We also find that the surface conductance is orders of magnitude higher than the bulk conductance close to the band edges for the unpassivated surface at room temperature. Therefore, we propose that the transport through these surface states may be the dominant conduction mechanisms in the recently reported scanning tunneling microscopy of silicon nanomembranes. We also calculate the zero bias conductance under flat-band condition for the weakly interacting dangling bond wires along and perpendicular to the dimer row direction and find similar trends. Extended H\"uckel theory is used for the electronic structure calculations, which is benchmarked with the GW approximation for Si and has been successfully applied to Si systems in past.
\end{abstract}


\maketitle

\begin{table}
\caption{Details of the model systems. The models consist of sixteen atomic layers shown in Fig. 1 to eliminate quantization effects and hence to obtain bulk Si band gap. The total number of Si atoms (excluding H atoms on top and back surface) per unit cell are also given. As shown in Fig. 1, a (2$\times$1) unit cell has two repeating unit cells in the [110] direction (perpendicular to dimer row) and one unit cells in the $[\overline{1}10]$ direction (along dimer row).}
\begin{ruledtabular}
\begin{tabular}{clcc}
Model & Unit cell surface & Unit cell\footnotemark[1] & Atoms\footnotemark[2]\\
\hline
I&H:Si(001)-(2$\times$1) & 2$\times$1 & 16\\
II&Si(001)-(2$\times$1)-AD\footnotemark[3] & 2$\times$1 & 16\\
III&Paired dangling bond wire\footnotemark[3] & 4$\times$1 & 32\\
IV&Si(001)-(2$\times$1)-AD\footnotemark[4] & 2$\times$1 & 16\\
V&Paired dangling bond wire\footnotemark[4] & 4$\times$1 & 32\\
\end{tabular}
\end{ruledtabular}
\footnotetext[1]{Multiples of 3.84 $\AA$ - lattice constant of bulk unit cell.}
\footnotetext[2]{Number of Si atoms only per unit cell.}
\footnotetext[3]{in $[\overline{1}10]$ direction.}
\footnotetext[4]{in [110] direction.}
\end{table}

\section{Introduction} Two dimensional (2D) electronic systems \cite{RMP82} have been of great interest. In metal-oxide-semiconductor (MOS) technology, the 2D electron and hole gas is the key for functioning of the modern integrated circuits. Also, in high-T$_c$ materials, the 2D cuprate planes are thought to be responsible for superconductivity. Graphene is another noteworthy example of 2D transport and has attracted tremendous interest recently.  

\begin{figure*}
\vspace{5.5in}
\hskip -3in\includegraphics{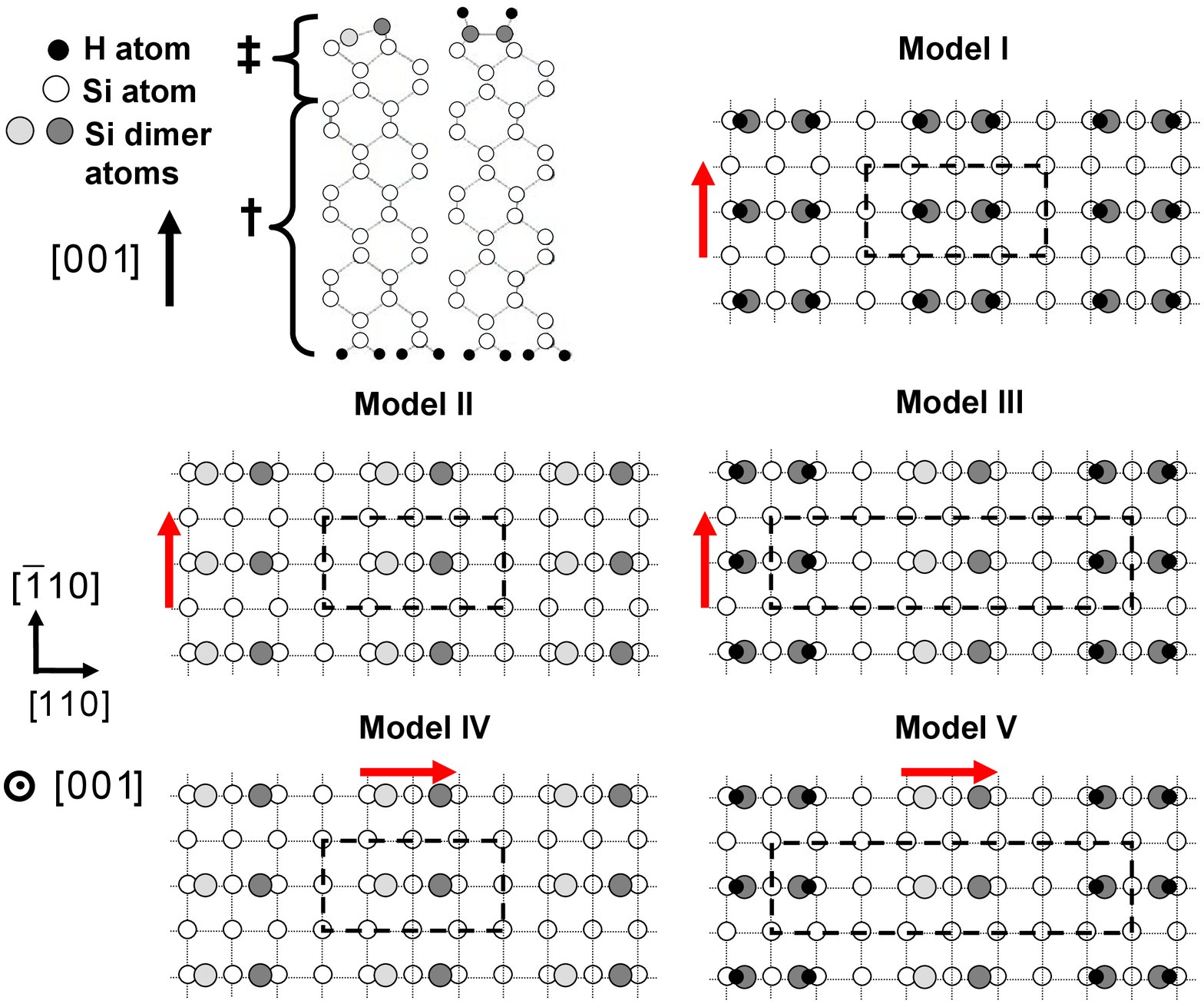}
\caption{(color online) Ball and stick models. Unpassivated paired asymmetric dimer (AD) surface and hydrogenated symmetric dimer surface. Top four layers, represented by $\ddag$, are relaxed due to surface reconstruction. Bottom twelve, represented by $\dag$, are bulk layers. The back surface is hydrogenated to eliminate any inadvertent DB induced states. The top Si atom of the AD surface is shaded as black and the bottom Si atom is shaded as gray. The atoms below the first mono-layer of Si surface are represented by white circles. The direction parallel to the dimer row is conveniently referred to as $[\overline{1}10]$, whereas the one perpendicular to the dimer row is labeled as [110]. Atomic coordinates for these structures are discussed and reported in Ref. \cite{Raza07}. Atomic visualization is done using GaussView \cite{GW03}.}
\end{figure*}

The electronic structure of the Si(001) surface has been a topic of study for decades. Although, some aspects of this particular surface are still controversial, overall it is a well understood surface \cite{SrivastavaBook, Raza07}. An unpassivated Si surface with paired asymmetric dimer (AD) reconstruction introduces two surface bands which have a 2D character \cite{SrivastavaBook, Raza07}. The band corresponding to the anti-bonding state ($\pi^*$ band) has acceptor states \cite{Sze}. Similarly, the band corresponding to the bonding state ($\pi$ band) has donor states \cite{Sze}. The surface state density is about $10^{15}\ cm^{-2}$ for $\pi^*$ and $\pi$ bands. Since the $\pi^*$ band is fully unoccupied and the $\pi$ band is fully occupied, they do not contribute to the transport in the absence of doping and surface band bending. Apart from this, the $\pi^*$ and $\pi$ states are localized on the bottom and top dimer atoms, respectively \cite{SrivastavaBook}. Moreover, the wavefunction is localized within 10 $\AA$ of surface \cite{Raza07,Raza06}, which is a very important length scale for the system under study. 

Furthermore, these two bands may start conducting with doping and/or surface band bending. In such a scenario, the transport through the $\pi^*$ and $\pi$ bands can be thought of as electron and hole transport, respectively. Recently, Zhang \textit{et al.} \cite{Zhang06} reported scanning tunneling microscopy (STM) of silicon nanomembranes with about $10^{15}\ cm^{-3}$ p-type doping. With the atomic resolution for this system in STM, they propose that surface doping enabled by thermal excitation of electrons in $\pi^*$ band results in holes in the valence band, which enables the hole conduction inside the valence band. Zhang \textit{et al.} \cite{Zhang06} place the $\pi$ band edge few tenths of eV below the valence band edge. We find that over the transverse Brillouin zone, the bottom of $\pi^*$ band is about 0.4 eV below the conduction band edge ($E_c$) and the top of the valence band is about 0.2 eV above the valence band edge ($E_v$). Additionally, our model predicts that the zero bias conductance through these surface states is at least three orders of magnitude higher than that of the bulk for the doping being used in STM \cite{Zhang06}. Therefore, we propose that the surface state transport alone may be the dominant conductance mechanism in silicon nanomembranes. 

We calculate transmission and zero bias conductance under flat-band condition for these surface states in directions parallel to and perpendicular to the dimer row referred to as $[\overline{1}10]$ and [110], respectively. We also calculate transmission and zero bias conductance under flat-band condition for paired dangling bond (DB) wires separated by hydrogenated DB wires in above-mentioned directions. To the best of our knowledge, these are the first atomistic transport calculations for the systems under consideration.

\section{Model Systems} In a previous study \cite{Raza06}, we report that an isolated unpaired DB on an otherwise perfectively hydrogenated Si(001) surface will only affect its neighboring Si atoms within 10 $\AA$. Same is the case for the paired DBs on the Si surface \cite{Raza07}. The $\pi^*$ and $\pi$ states, respectively are localized on the bottom and top Si atoms in the paired AD. In the [$\overline{1}$10] direction, the Si dimer atoms, on which $\pi^*$ and $\pi$ states are present, are only 3.84 $\AA$ apart, while in [110] direction, about 7.68 $\AA$ apart. The hybridization anisotropy in [$\overline{1}$10] and [110] directions results in different band-widths \cite{Raza07}. The extent of the wavefunction overlap also affects the transport in these directions. 

\begin{figure}
\vspace{3.5in}
\hskip -2in\includegraphics{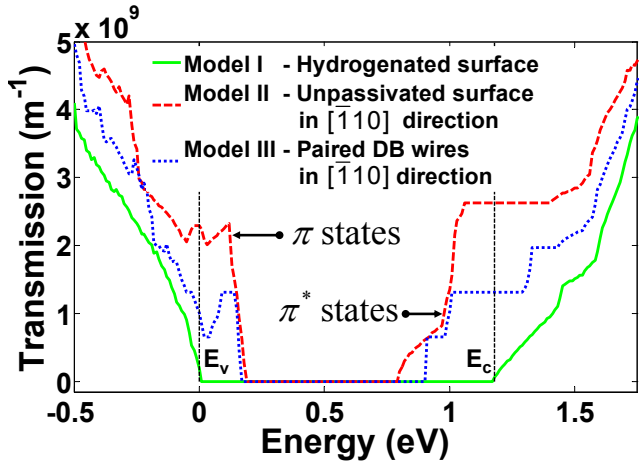}
\caption{(color online) Transmission calculations for Model I, II and III. Hydrogenated Si(001) surface (model I) gives bulk Si band gap of 1.18 eV. Unpassivated Si(001) surface (model II) gives large transmission in the dangling bond row direction ($[\overline{1}10]$) for both the $\pi^*$ and the $\pi$ states because the dimer atoms on which these states are localized are 3.84 $\AA$ apart. Transmission through the paired dangling bond (model III) wire along the dangling bond row direction, is about half of that in model II due to half of the total dangling bond states per unit transverse length.}
\end{figure}

In this paper, we consider five model systems for the transport calculations described in Table I and Fig. 1. The transport direction is shown by arrow and the unit cell used for each model is shown by the dashed line. Model I is a hydrogenated surface and is expected to have bulk Si band gap. We calculate transport for this model in [$\overline{1}$10] and use it as a reference for the other models. Model II and IV are unpassivated surfaces and the transport is calculated in the [$\overline{1}$10] and the [110] directions, respectively. Model III and V are paired DB wires separated by hydrogenated wires and transport is calculated in [$\overline{1}$10] and [110] directions, respectively. The size of the unit cell used and number of atoms per unit cell are also given in Table I. Since DBs interact within 10 $\AA$, enough neighboring unit cells are included to calculate Hamiltonian (H) and overlap (S) matrices for the E($\overrightarrow{k}$) calculations.

\begin{figure}
\vspace{3.5in}
\hskip -2in\includegraphics{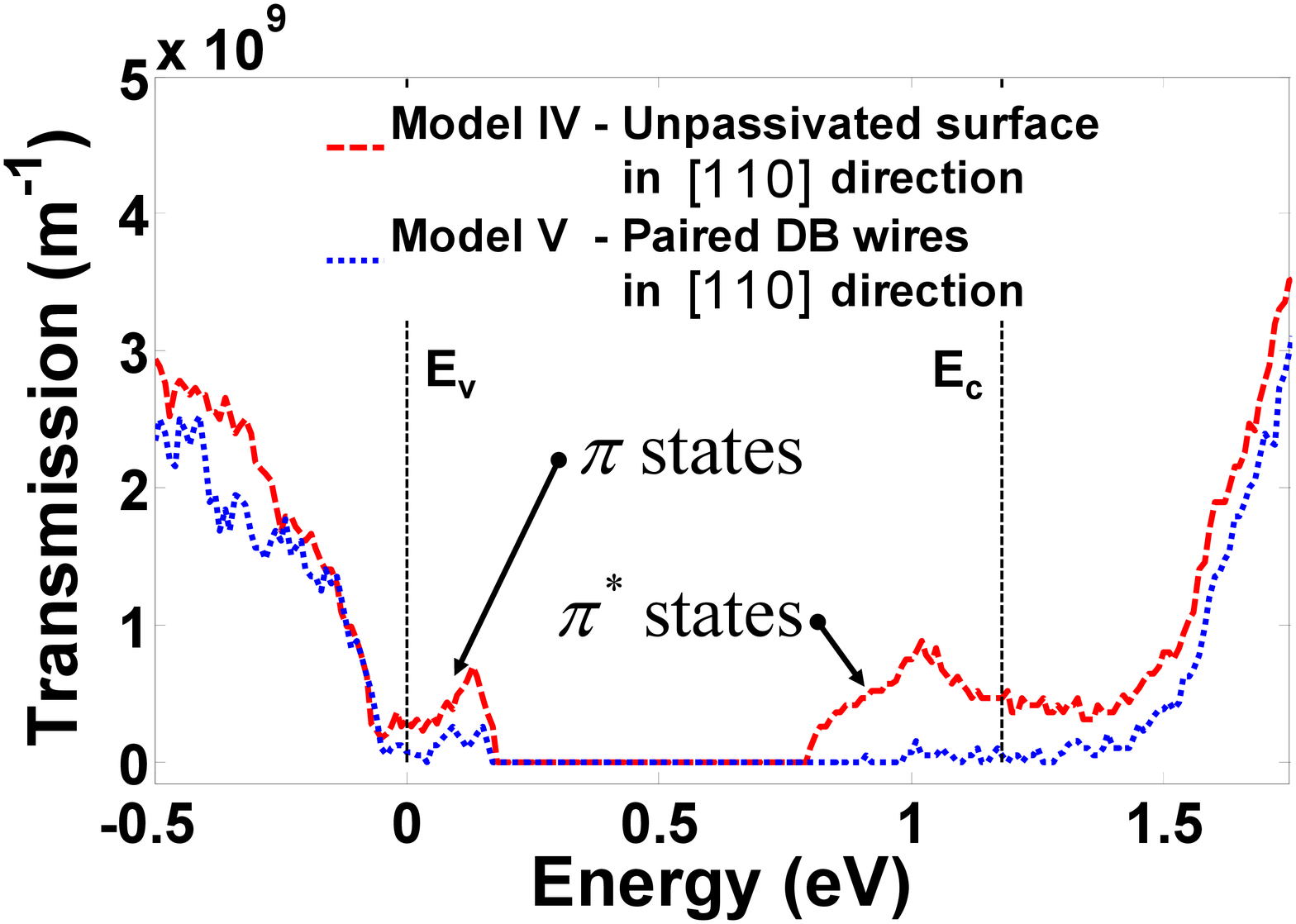}
\caption{(color online) Transmission calculations for models IV and V. For model IV, transmission through Si surface states in the direction perpendicular to dimer row direction ($[110]$) is smaller than the one along dimer row direction (model II) due to the reduced hybridization between surface states - DB atoms being 7.68 $\AA$ apart. Conclusion for the DB wire, for which the transmission is calculated perpendicular to dimer row direction (model V) is the same. }
\end{figure}

\section{Theoretical Calculations} We use extended H\"uckel theory (EHT) for the electronic structure calculations as in Ref. \cite{Raza07}. EHT prescribes a semi-empirical tight-binding procedure using a Slater type orbital non-orthogonal basis set. For Si, EHT is benchmarked \cite{Cerda00} with GW approximation and thus gives correct band structure features, such as band offsets and dispersions. Transferable EHT parameters used in this paper are taken from Ref. \cite{Cerda00} and are summarized in Table II. We use one orbital (1s) basis set for the H atom, whereas a nine-orbital (3s, 3p and 3d) basis set is used for the Si atom. 

\begin{figure*}
\vspace{5in}
\hskip -3in\includegraphics{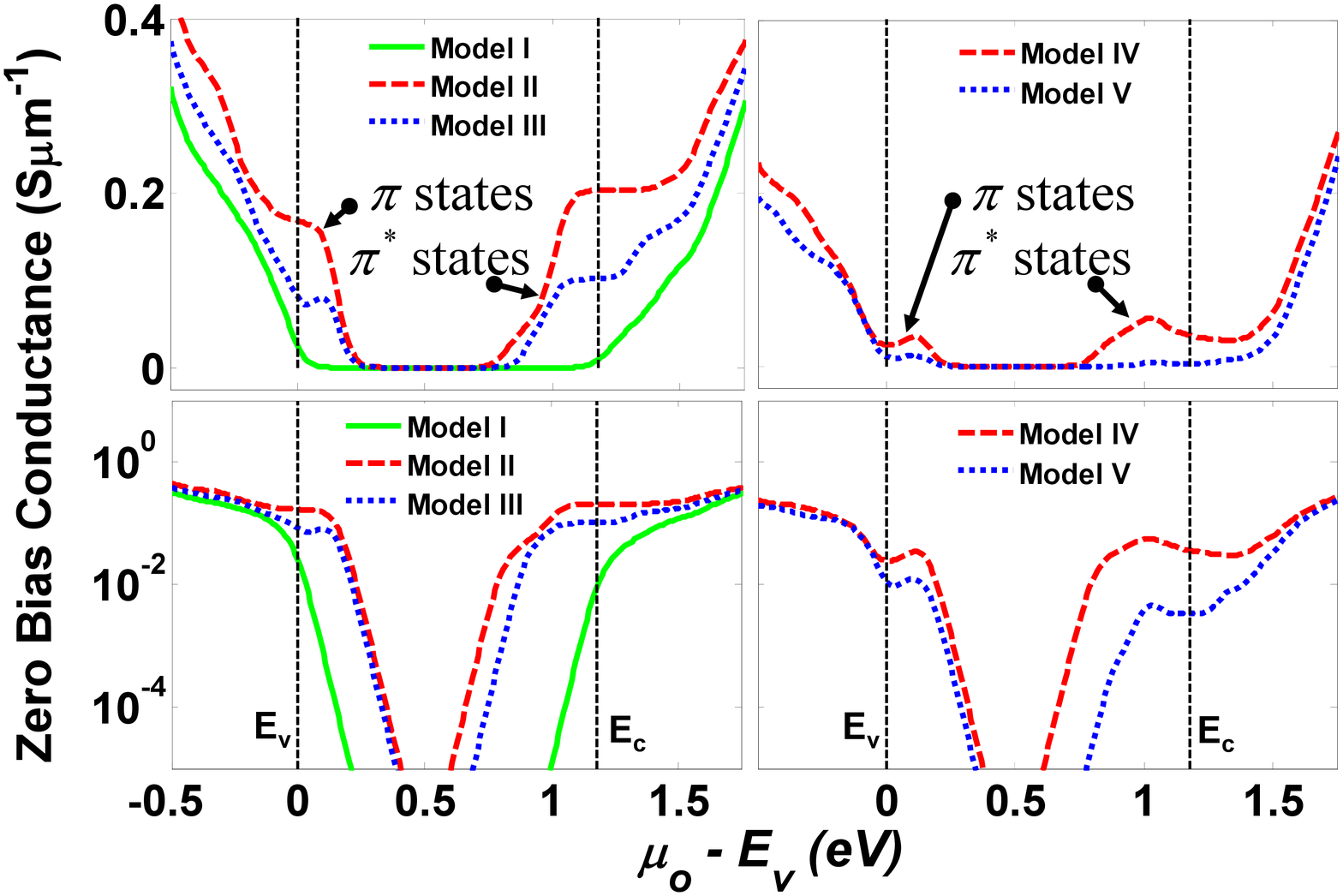}
\caption{(color online) Zero bias conductance under flat-band condition. For model I (hydrogenated surface with transport in the dimer row direction), the conductance decreases exponentially inside the band gap due to Fermi tail. Close to the band edge, the zero bias conductance is orders of magnitude higher for model II (unpassivated surface with transport in dimer row direction) than that of model I. The zero bias conductance conductance for model III (DB wire with transport in dimer row direction) follows the same trend as that of model II with some quantitative differences. The zero bias conductance for the $\pi^*$ states of model IV (unpassivated surface with transport perpendicular to the dimer row direction) is two orders of magnitude higher and less than one order of magnitude higher than that of model V (DB wire with transport perpendicular to the dimer row direction) inside the band gap.}
\end{figure*}

In order to calculate the transport properties for a 2D channel, we first calculate the E(k) diagrams over the transverse Brillouin zone for each wavevector in the transverse direction ($k_t$) by transforming the real space Hamiltonian (H) and overlap (S) matrices to reciprocal ($\overrightarrow{k}$) space:
\begin{eqnarray}H(\overrightarrow{k})=\sum_{m=1}^N{H_{mn}e^{i\overrightarrow{k}.(\overrightarrow{d_m}-\overrightarrow{d_n})}}\end{eqnarray}
\begin{eqnarray}S(\overrightarrow{k})=\sum_{m=1}^N{S_{mn}e^{i\overrightarrow{k}.(\overrightarrow{d_m}-\overrightarrow{d_n})}}\end{eqnarray}
where $\overrightarrow{k}=(k_t,k)$ - k is the wavevector in the transport direction. For each transverse wavevector ($k_t$), the system thus becomes one-dimensional (1D) and hence its transmission is independent of the dispersion. This enables transmission to be calculated numerically by counting band-crossings at a particular energy for each $k_t$.  Finally transmission per unit length is calculated by summation over transverse $k_t$ as follows:
\begin{eqnarray}T(E)=\frac{1}{L}\sum_{k_t}{\tilde{T}(E_l,k_t)=\frac{1}{2\pi}\int{dk_t}\tilde{T}(E_l,k_t)}\end{eqnarray}
Additionally, the zero bias conductance ($G_o$) under flat-band condition at a finite temperature is calculated as (see appendix):
\begin{eqnarray}G_o= \frac{2q^2}{h}\frac{1}{kT}\int{ dE\ T(E)\  \frac{e^{(E-\mu_o)/kT}}{[1+e^{(E-\mu_o)/kT}]^2}}\end{eqnarray}

\section{Discussion of Results} We report the transmission per unit transverse length through model I in Fig. 2, which shows clean bulk band gap of 1.18eV as expected due to the hydrogen passivation and is used as a reference for the other calculations. For the AD surface (model II), transmission through the $\pi^*$ and $\pi$ states is shown in Fig. 2 along the $[\overline{1}10]$ direction. The atoms are 3.84 $\AA$ apart on which $\pi^*$ and $\pi$ states are localized. Since wavefunctions are extended up to 10 $\AA$, hybridization is strong and hence the transmission is large. The transmission through the $\pi^*$ state starts increasing about 0.5 eV below the conduction band edge. The maximum transmission through the $\pi^*$ state is about $3\times10^{9}\ /m$, which gives a zero bias conductance of about $0.2 \ S/\mu m$ according to Eq. 4 and is shown in Fig. 4. Additionally, the transmission through the $\pi$ state is about 0.2 eV above the valence band edge. For the model III, the transmission is again large because atoms corresponding to $\pi^*$ and $\pi$ states are  3.84 $\AA$ apart. But since the overall density of dangling bond is half as that of model II, the transmission is almost half for both the states. However, there are some differences in characteristic features due to varying band offsets at various transverse wavevectors ($k_t$) for models II and III.

\begin{table}
\caption{EHT parameters used for H and Si atoms. $K_{EHT}=2.3$}
\begin{ruledtabular}
\begin{tabular}{lccccc}
Orbital & $E_{on-site}(eV)$ & $C_1$ & $C_2$ & $\xi_1$ ($\AA^{-1}$) & $\xi_2$ ($\AA^{-1}$)\\
\hline
H:1s  & -16.15180 & 0.53558 &         & 0.88560 &         \\
Si:3s & -18.10264 & 0.70366 &         & 1.83611 &         \\
Si:3p & -11.25298 & 0.02770 & 0.98313 & 0.78901 & 1.70988 \\
Si:3d & -5.347060 & 0.68383 & 0.46950 & 0.68292 & 1.72102 \\
\end{tabular}
\end{ruledtabular}
\end{table}

Models IV and V are the same as models II and III, but the transmission is calculated in $[110]$ direction. As shown in Fig. 3, the overall transmission is smaller even inside the conduction and the valence bands, because Si atoms in the top four surface layers are further apart due to asymmetric reconstruction. Apart from this, for model IV, since atoms corresponding to $\pi^*$ and $\pi$ states are now 7.68 $\AA$ apart, there is small hybridization between atoms on which $\pi^*$ and $\pi$ bands are localized and hence surface state hopping is quenched. The maximum transmission inside band gap for model IV is about $0.75\times10^{9}\ /m$, which gives a conductance of about $0.05\ S/m$. For model V, the unpassivated bottom and top dimer atoms in the transport direction are about 15.36 $\AA$ apart, which results in very small hybridization of the $\pi^*$ and the $\pi$ states. Therefore, the transmission through these states for model V is even smaller. 

We show the calculated zero bias conductance under flat-band condition at room temperature on a linear and a logarithmic scale in Fig. 4. The conductance for model II and III is orders of magnitude higher than the model I inside the band gap. For model IV, the conductance is about two orders of magnitude higher for the $\pi^*$ states and less than one order of magnitude higher for the $\pi$ states than those of model V. Furthermore, the zero bias conductance decreases exponentially inside the band gap due to the Fermi tail. For the experimental conditions used in Ref. \cite{Zhang06}, the band bending is small and hence based on our flat-band zero bias conductance calculations, we propose that the surface state hoping is the dominant conduction mechanism in the recently reported STM of lightly doped silicon nanomembranes \cite{Zhang06}. 

Since the transport is coherent and the channel is in the ballistic regime, the transmission is independent of the channel length. Hence, the results reported provide an upper limit of transmission and conductance for these surface states. In this study, effects of gate voltage and realistic contacts are ignored. In Ref. \cite{SiSS-archive}, we provide the basic formalism to perform these calculations. Furthermore, we leave the role of defects, the electrostatic effects of the dopant atoms and dephasing due to electron-phonon scattering for future work. Interestingly, it has been suggested that atomic steps on Si(111) surface can decrease conduction by two orders of magnitude \cite{Petersen00}.

\section{Conclusions} 

We have studied the transport through Si(001)-(2$\times$1) surface states with asymmetric dimer reconstruction in the dimer row and its perpendicular direction. We find significant transmission in the dimer row direction due to strong hybridization between dangling bonds 3.84 $\AA$ apart. However, in the direction perpendicular to the dimer row, bottom and top dimer atoms are about 7.68 $\AA$ apart, on which $\pi^*$ and $\pi$ states are localized, their hybridization is weak. We find similar trends for the paired dangling bond wires separated by hydrogenated wires. Furthermore, the zero bias conductance under flat-band condition at room temperature around the band edges is at least three orders of magnitude higher for these surface band in comparison with that of passivated surface. Therefore, we propose that conduction through the surface states may be the dominant conduction mechanism in the recently reported scanning tunneling microscopy results of silicon nanomembranes on insulator.

\section*{Appendix}

We derive Eq. 4 as follows. For a finite voltage V, the difference of the Fermi function is given as:

\begin{eqnarray}f_1-f_2=\frac{1}{1+e^{(E-\mu_o)/kT)}}-\frac{1}{1+e^{(E-\mu_o+qV)/kT)}}\nonumber\end{eqnarray}
\begin{eqnarray}f_1-f_2=\frac{e^{(E-\mu_o)/kT}[e^{qV/kT}-1]}{[1+e^{(E-\mu_o)/kT}][1+e^{(E-\mu_o+eV)/kT}]}\nonumber\end{eqnarray}

For $|qV| \ll kT$, using Taylor series expansion and retaining the first term for $e^{qV/kT}$, it can be shown that:

\begin{eqnarray}f_1-f_2=\frac{qV}{kT}\frac{e^{(E-\mu_o)/kT}}{[1+e^{(E-\mu_o)/kT}]^2}\nonumber\end{eqnarray}

Starting with Landa\"uer's approach, 

\begin{eqnarray}I=\frac{2q}{h}\int{dE\ T(E) (f_1-f_2)}\nonumber\end{eqnarray}

it can be shown that zero bias conductance at a finite temperature is given as:

\begin{eqnarray}G_o=\frac{I}{V}=\frac{2q^2}{h}\frac{1}{kT}\int{dE\ T(E)\ \frac{e^{(E-\mu_o)/kT}}{[1+e^{(E-\mu_o)/kT}]^2}}\nonumber\end{eqnarray}

\section*{Acknowledgments}

We thank M. G. Lagally and A. Alam for useful discussions. The work is supported by National Science Foundation (NSF) and by Nanoelectronics Research Institute (NRI) through Center for Nanoscale Systems (CNS) at Cornell University.

\end{document}